\documentclass{appolb}
\usepackage{graphicx}
\begin{document}
\title{Meson cloud effects in nucleon resonances at low and 
intermediate energies%
\thanks{Presented at EEF70, Workshop on Unquenched Hadron Spectroscopy, 
1-5 Sept. 2014, Coimbra}%
}
\author{Bojan Golli
\address{Faculty of Education,
              University of Ljubljana,
              1000 Ljubljana, Slovenia
 and
Jo\v{z}ef Stefan Institute, 
              1000 Ljubljana, Slovenia}
}
\maketitle

\begin{abstract}
We review some results obtained by the Coimbra-Ljubljana
collaboration that reveal the important role of the pion cloud
in the low- and intermediate-energy nucleon and delta resonances.
\end{abstract}
\PACS{{}25.80.Ek, 12.39.Fe, 13.60.Le, 14.20.Gk}

\section{Introduction}
The notion that the pion cloud plays an important role in the 
formation of nucleon resonances was first anticipated by 
Jo\~ao da Provid\^encia and Jo\~ao Urbano in 1978 \cite{JdaP78}.
In their work the nucleon resonances arise as excitations of 
the pion cloud around the bare nucleon. 
The classical pion field was interpreted as a coherent state 
of pions and the resonances with good angular momentum and 
isospin were obtained by Peierls-Yoccoz projection.
This approach is similar to the so called model of 
{\em dynamical generation\/} of resonances. 
On the other hand in the {\em quark model\/} the resonances 
emerge as excitations of the three-quark core.
In early eighties we started a long term collaboration 
between the Coimbra group (J. da Provid\^encia, M. Fiolhais, 
P. Alberto, and L. Amoreira) and the Ljubljana group 
(M. Rosina, S. \v{S}irca, M. \v{C}ibej and B.~G.).
In our work we have combined both approaches assuming 
a superposition of different excitations of the quark core 
surrounded by meson clouds.
In such a picture, some resonances can be described as
almost pure single-quark excitations of the core while
for other resonances the main mechanism is the excitation 
of the meson cloud.
How can we determine which mechanism dominates in a given 
resonance?
If we investigate just meson scattering in the resonance region
we may not reliable determine the nature of the resonance 
since by a suitable modification of the coupling constants 
and the interaction range it possible to reproduce reasonably 
well the elastic and inelastic amplitudes almost in any model.
We claim that studying meson electro-production processes, 
in particular the $Q^2$-dependence of helicity amplitudes, 
is a much more severe test to analyze the structure of 
various resonances.

In the next sections we present some examples in which we 
have established the important role of the meson cloud. 
We have considered different chiral quark models, the linear 
$\sigma$-model with quarks, the chromodielectric model in 
which a dynamical confinement is generated by an additional 
scalar field, and the cloudy bag model, which turns out to 
be suitable for the description of higher resonances.

\section{The $\Delta$(1232) resonance}

The $\Delta$(1232) was the first resonance that we tackled.
We assumed a superposition of the bare $\Delta$ core with 
three quarks in the 1s orbit with the spins and isospins 
coupled to 3/2, and the pion cloud around the bare nucleon 
and the bare $\Delta$ core.
Using Peierls-Yoccoz projection we maintained the correct 
spin and isospin of the composite state.
We calculated the electromagnetic transition amplitudes as
a function of the photon virtuality $Q^2$.
We found that the processes in which the photon interacts
with the pion cloud importantly contribute to the helicity
amplitude \cite{delta96}.
For the leading magnetic dipole contribution (M1) we could 
have anticipated such a result from our experience with the 
nucleon magnetic moments.
On the other hand, the quadrupole excitations, the transverse 
E2 and the Coulomb C2, are not allowed in the nucleon, but 
are present in the transition $N$ to $\Delta$ amplitude.
We have found relatively large quadrupole amplitudes 
originating from the $p$-wave pion flip of its third component 
of the angular momentum. 
\begin{figure}[htb]
\centerline{%
\includegraphics[width=4.5cm]{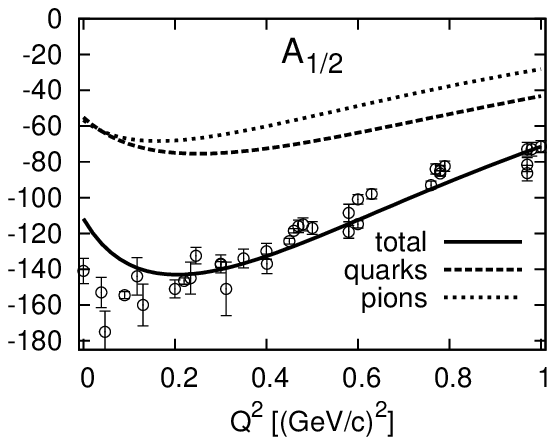}\hspace{-4mm}
\includegraphics[width=4.5cm]{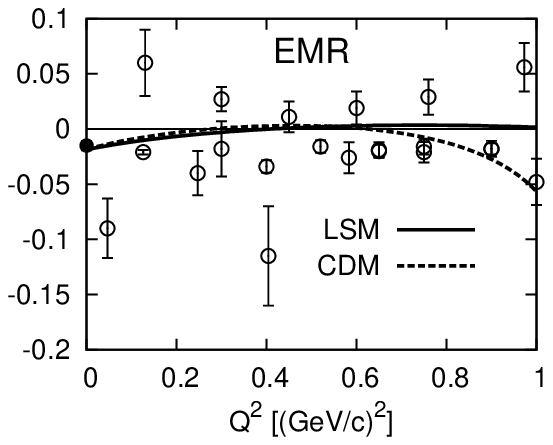}\hspace{-4mm}
\includegraphics[width=4.5cm]{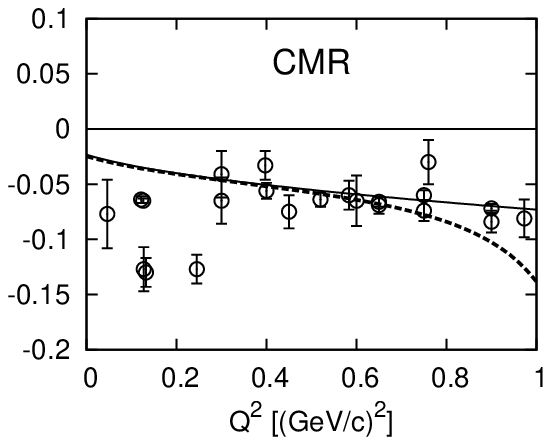}}
\caption{The separate quark and pion contributions
to the helicity amplitude $A_{1/2}$ in units of 
$10^{-3}~\mathrm{GeV}^{-1/2}$ calculated in the linear 
$\sigma$-model (left panel), the ratio $E2/M1$ (middle panel)
and $C2/M1$ (right panel) calculated in the linear 
$\sigma$-model and in the cromodielectric model.}
\label{Fig:Delta}
\end{figure}
In a quark model without pions such a strong contribution
could only be possible if we assumed an unrealistically 
strong admixture of the $d$-wave quark configuration.
To the best of our knowledge, we were the first to calculate
the $Q^2$ behavior of the E2 and C2 amplitude.
We showed that the different $Q^2$ behavior of these 
amplitudes (equal for a point-like source) originates from 
the long-range part of $N$ and $\Delta$ wave function 
which is dominated by the pion tail.
Since then, our result has been rediscovered in different 
approaches and supported by new more precise measurements.

\section{The multi-channel approach}
While for the $\Delta$(1232) resonance the inelastic channels
can be ignored, they have to be included if we want to study
higher resonance.
We have developed an approach for computing the multi-channel 
$K$ matrix which includes many-body quasi-bound quark states 
in the scattering formalism \cite{EPJ2008}.
In this approach the channel state corresponding to the 
scattering baryon and meson is (up to the normalization factor) 
given by
\begin{equation}
|\Psi^{MB}_{JI}\rangle = 
        a^\dagger(k)|\Psi_B\rangle
        + c_{\mathcal{R}}
          |\Phi_{\mathcal{R}}\rangle
        + \sum_{M'B'}
          \int {{\rm d} k'\;
          \chi^{M'B'\,MB}(k',k)\over\omega_k+E_{B'}-W}\,
           a^\dagger(k')|\Psi_{B'}\rangle.
\label{PsiH}
\end{equation}
The first term represents the free meson (M) and the baryon (B)
and defines the channel, the next term is a bare three-quark
state, and the third term describes meson clouds around 
different isobars.
Here $W$ is the invariant energy, $J$ and $I$ are the angular 
momentum and isospin of the meson-baryon system.
The integration is taken in the principal-value sense.
The multichannel $K$ matrix can then be expressed as
\begin{equation}
 K_{M'B'\,MB}(k,k') =  -\pi\sqrt{\omega E_{B} \over k W}\;
   \langle\Psi^{MB}||V_{M'}(k')||\Psi_{B'}\rangle\,,
\label{defK}
\end{equation}
where $V_{M'}(k')$ stands for the quark-meson vertex
to be computed in the underlying quark model.
Above the $MB$ threshold the meson amplitudes $\chi$ 
in (\ref{PsiH}) are proportional to the $K$ matrix and 
are computed using the Lippmann-Schwin\-ger equation.
Finally, the scattering $T$ matrix is obtained by solving
the Heitler equation $T=K + {\rm i}TK$.

The meson electro-production amplitudes can be calculated
in the same framework by including the $\gamma N$ channel.
Close to a resonance, the amplitude can be cast in the form
$$
{\mathcal{M}_{MB\gamma N}^\mathrm{res}}  =
\sqrt{\omega_\gamma E_N^\gamma \over \omega_\pi E_N }\,
{\xi\over\pi{\cal V}_{N\mathcal{R}}^\pi}\,
  \langle\widehat{\Psi}_{\mathcal{R}}|{V}_\gamma
                |\Psi_N\rangle\, {T_{MB\,\pi N}} \>,
$$
where the photon vertex $V_\gamma$ acquire contributions from 
quarks and pions, $\xi$ is the spin-isospin factor depending on 
the multipole and the spin and isospin of the outgoing hadrons.
Here $\langle\widehat{\Psi}_{\mathcal{R}}|{V}_\gamma|\Psi_N\rangle$
is the helicity amplitude; the resonance state
$\widehat{\Psi}_{\mathcal{R}}$ is extracted from the residua in the 
second and the third term in (\ref{PsiH}) at the resonance pole.

\section{The Roper resonance}
As the underlying quark model we have taken the cloudy bag 
model, primarily because of its simplicity.
In the first step we considered only two inelastic channels,
the $\pi\Delta$ channel, and -- assuming that the two-pion decay
proceeds mainly through the $\sigma$-meson -- the $\sigma N$ 
channel.
We have further assumed that the bare quark state in (\ref{PsiH})
is a mixture of the bare nucleon state and the bare Roper state 
in which one quark is excited to the 2s orbit:
$\Phi_{\mathcal{R}} = \sin\theta (1s)^3 + \cos\theta (1s)^2(2s)^1$.
\begin{figure}[htb]
\centerline{%
\includegraphics[width=5.3cm]{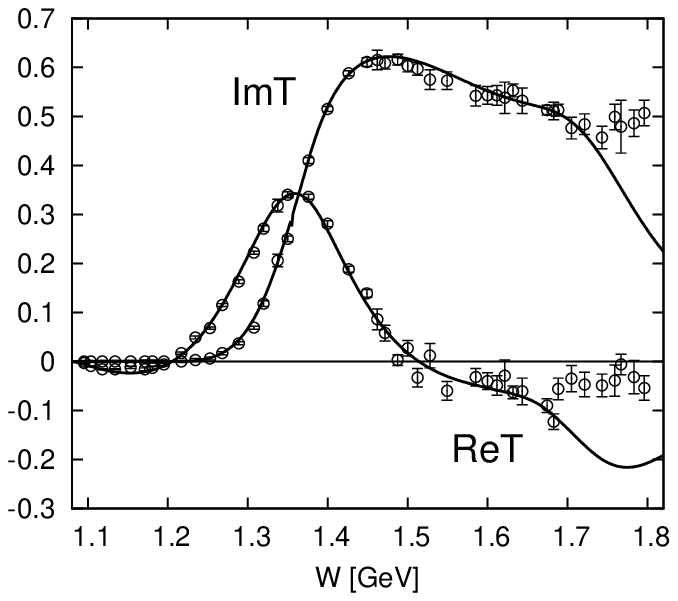}%
\vbox{\hbox{\includegraphics[width=3cm]{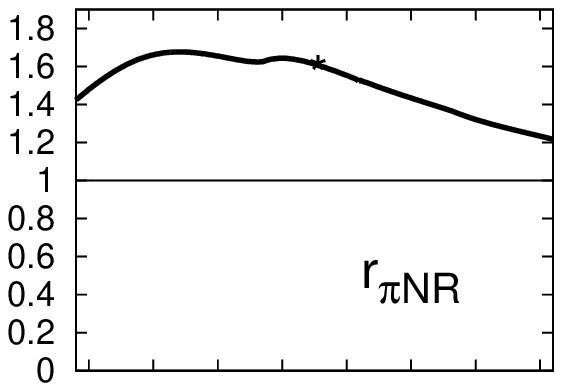}}%
   \vspace{-6pt}
   \hbox{\hspace{1pt}\includegraphics[width=3cm]{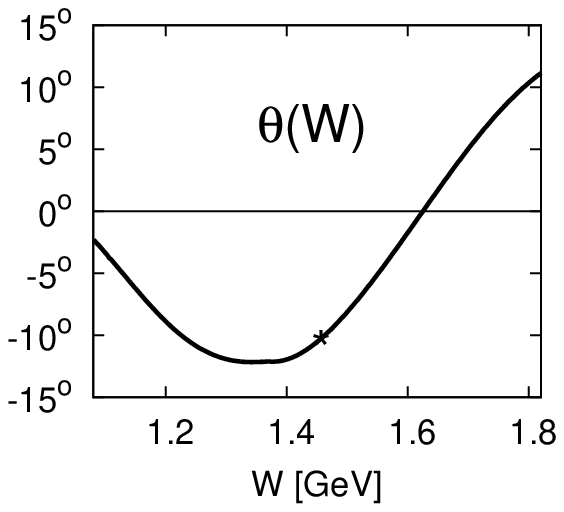}}}%
\includegraphics[width=5.2cm]{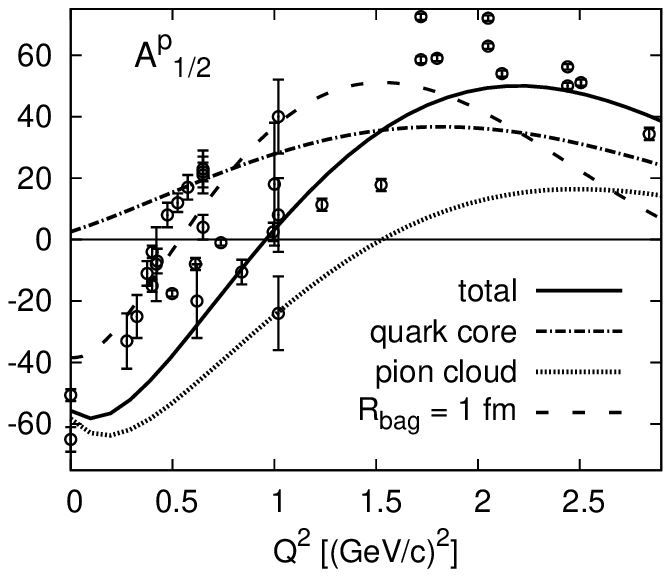}}
\caption{The real and the imaginary part of the $T$ matrix in 
the P11 partial wave (left panel), the renormalized $\pi NR$ 
coupling constant and the mixing angle $\theta$ as a function 
of $W$ (middle panels), the quark and the pion contribution 
to the helicity amplitude $A_{1/2}^p$ in units of 
$10^{-3}~\mathrm{GeV}^{-1/2}$ calculated in the cloudy bag model 
using the bag radius of 0.83~fm (right panel).}
\label{Fig:Roper}
\end{figure}
Using only the three channels we were able to reproduce the
experimental behavior of the scattering amplitudes up to
$W\sim 1600$~MeV; to reproduce the amplitudes at somewhat 
higher $W$ we included the second Roper resonance.
In quark models, the $\pi NR$ coupling constant turns out
to be too weak to reproduce the large resonance width, 
mostly due to the orthogonality of the 1s and 2s orbits.
In our approach, this coupling is strongly enhanced through 
the pion loops as well as through the mixing of the 
two bare configurations as shown in Fig.~\ref{Fig:Roper}.

The role of the pion cloud is most strongly pronounced
in the calculation of the helicity amplitude \cite{EPJ2009}.
Here the pion cloud plays an important role through the 
$\gamma NR$ vertex renormalization as well as through the 
direct coupling of the photon to the pion.
In the region of low $Q^2$ the quark contribution is small 
and positive, while the pion contribution and the vertex 
corrections due to meson loops are large and negative.
At intermediate $Q^2$, these two effects are responsible 
for the zero crossing of the amplitude.  
Using the bag radius of 1~fm instead of our standard choice
of 0.83~fm we are able to reproduce the popular value of
0.5~(GeV/$c$)$^2$ for the crossing point.
At higher $Q^2$ the quark core takes over, rendering the 
amplitude positive. 

Let us mention that the sign of the helicity amplitude 
calculated in a model is ambiguous since the resonance 
production is a non-observable process.
By convention, the sign is determined by the sign of the
resonance decay within the same model.
It is therefore not possible to compare different calculations
of the helicity amplitudes if the sign of the decay amplitude
is not provided.

\section{The negative parity resonances}
Here the pion cloud consists predominantly of the $s$- and 
$d$-wave pions which are less strongly coupled to quarks 
than the $p$-wave pions in the case of positive parity 
resonances.
Nonetheless we have found some interesting phenomena that
reveal the importance of the meson cloud also in this case.

In the S11 partial wave there are two relatively close-lying 
resonances, the $N(1535)$ and $N(1650)$.
In the quark model these resonances appear as mixtures 
of two \underline{70}-plet states with spin 1/2 and 3/2,
i.e.: $|\Psi(1535)\rangle = \cos\theta|{}^2{\bf 8}_{1/2}\rangle 
                           -\sin\theta|{}^4{\bf 8}_{1/2}\rangle$
and $|\Psi(1650)\rangle = \sin\theta|{}^2{\bf 8}_{1/2}\rangle 
                         +\cos\theta|{}^4{\bf 8}_{1/2}\rangle$.
\begin{figure}[htb]
\centerline{%
\includegraphics[width=6cm]{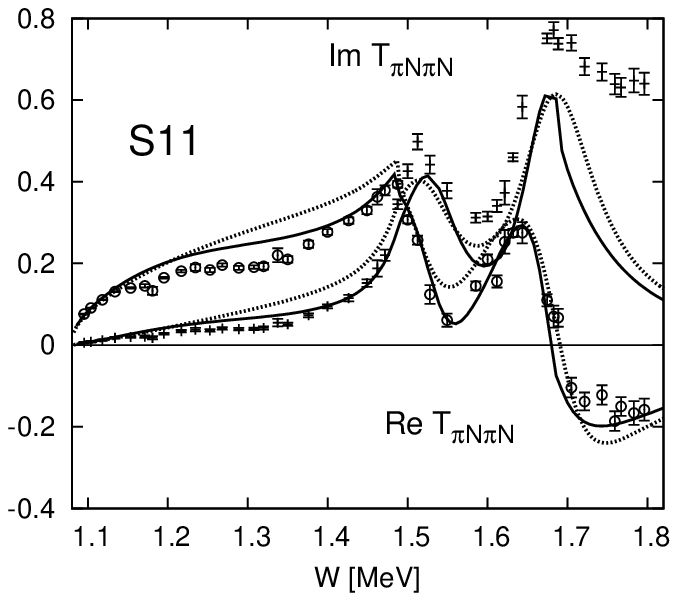}%
\vbox{\hbox{\includegraphics[width=3.45cm]{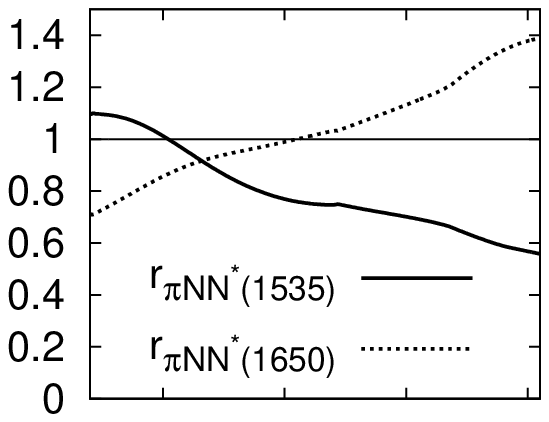}}
      \vspace{-6pt}
      \hbox{\hspace{5.5pt}\includegraphics[width=3.2cm]{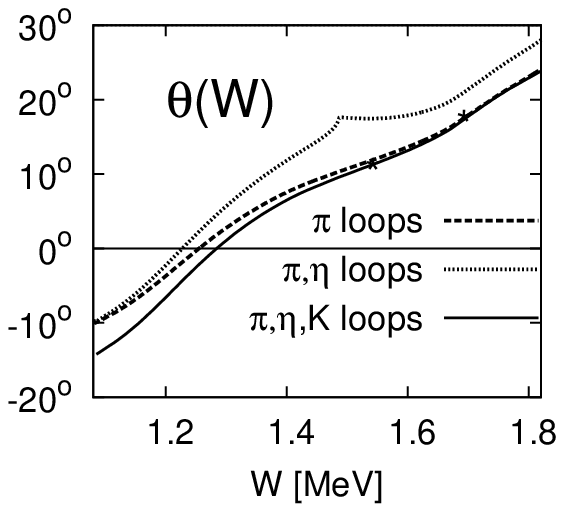}}}
}
\caption{The scattering $T$ matrix in the S11partial wave (left panel),
the renormalized $\pi NN^*$ coupling constants for the two resonance
and the mixing angle between the corresponding bare states
as a function of $W$ (middle panels).}
\label{Fig:S11}
\end{figure}
It turns out that the mixing is such that only the lower state 
couples strongly to the $\eta N$ channel while the coupling to 
this channel is almost absent for the higher state.
In our calculation \cite{EPJ2011} we took a fixed angle
of $\theta\approx 30^\circ$.
A more elaborate calculation, including pion, $\eta$-meson and 
kaon loops shows that the mixing angle is generated entirely
by the meson cloud effects and has a strong $W$ dependence.
As a consequence, it improves the behavior of the scattering
amplitudes at smaller values of $W$ (solid lines in Fig.~\ref{Fig:S11})
over the calculation with the fixed value of $\theta$ (dotted lines).

The pion cloud contribution to the helicity amplitudes is 
generally weak, of the order of $~10$~\%, except in the case 
of the $\Delta(1700)\; {3/2}^{-}$ resonance.
Here we have found an almost equal contribution of the quark core
and the pion cloud, quite a similar situation as in the case of 
its positive parity counterpart, the $\Delta(1232)$ \cite{EPJ2013}.
\begin{figure}[htb]
\centerline{%
\includegraphics[width=5.5cm]{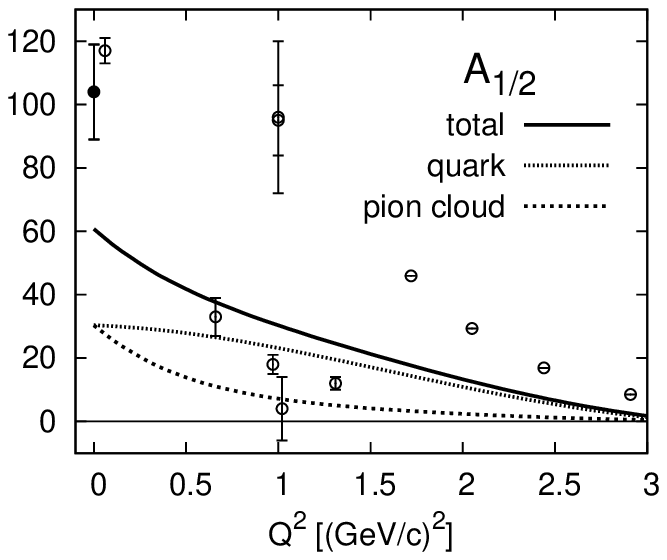}
\includegraphics[width=5.5cm]{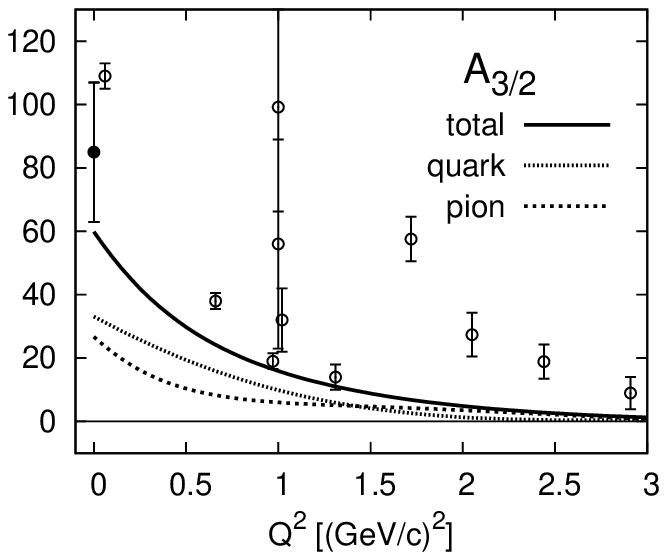}}
\caption{The separate quark and pion contributions to the helicity 
amplitudes $A_{1/2}$ and $A_{3/2}$ for the $\Delta(1700)$ resonance.}
\label{Fig:D33}
\end{figure}

\bigskip 

\noindent
The author would like to acknowledge the hospitality 
and the stimulating atmosphere he enjoyed during the workshop 
at the University of Coimbra.

\end{document}